\documentclass[conference]{IEEEtran}

\usepackage{cite}
\usepackage{amsmath,amssymb,amsfonts}
\usepackage{textcomp}
\usepackage{xcolor}
\usepackage{booktabs}
\usepackage{tabularx}
\usepackage{array}
\usepackage{multirow}
\usepackage[hyphens]{url}
\usepackage{tikz}
\usetikzlibrary{arrows.meta,fit,positioning,shapes.geometric}
\usepackage{listings}
\usepackage{hyperref}
\usepackage{amsthm}

\theoremstyle{definition}
\newtheorem{example}{Example}
\definecolor{diffminus}{RGB}{180,45,45}
\definecolor{diffplus}{RGB}{35,130,75}

\lstdefinestyle{diffsnippet}{
  basicstyle=\ttfamily\scriptsize,
  breaklines=true,
    columns=fullflexible,
    frame=single,
    framesep=2pt,
    morecomment=[l][\color{diffminus}]{-},
    morecomment=[l][\color{diffplus}]{+},
    xleftmargin=4pt,
    xrightmargin=4pt,
    aboveskip=\medskipamount,
    belowskip=\medskipamount,
}

\definecolor{scrollbg}{RGB}{252,245,224}
\definecolor{scrollframe}{RGB}{120,82,45}
\definecolor{scrolllinenumber}{RGB}{130,100,70}
\definecolor{scrollkeyword}{RGB}{22,86,140}
\definecolor{scrollcomment}{RGB}{74,117,76}
\definecolor{scrollstring}{RGB}{156,92,24}
\definecolor{scrolltext}{RGB}{45,35,25}

\lstdefinestyle{casestudyscroll}{
    backgroundcolor=\color{scrollbg},
    basicstyle=\ttfamily\scriptsize\color{scrolltext},
    commentstyle=\itshape\color{scrollcomment},
    keywordstyle=\bfseries\color{scrollkeyword},
    stringstyle=\color{scrollstring},
    numberstyle=\tiny\color{scrolllinenumber},
    numbers=left,
    numbersep=8pt,
    frame=single,
    framerule=0.9pt,
    rulecolor=\color{scrollframe},
    xleftmargin=8pt,
    xrightmargin=6pt,
    framesep=6pt,
    tabsize=2,
}

\newcommand{\depname}{DepBench}
\newcommand{\benchmarkname}{\textsc{\depname}}

\newcommand{\code}[1]{\emph{#1}}

\newcommand{\rqanswer}[1]{%
  \par\smallskip
  \noindent\begingroup
  \setlength{\fboxsep}{5pt}%
  \colorbox{black!8}{%
    \parbox{\dimexpr\linewidth-2\fboxsep\relax}{\normalsize
      \textbf{Answer.} #1}%
  }%
  \endgroup
  \par\smallskip
}

\def\BibTeX{{\rm B\kern-.05em{\sc i\kern-.025em b}\kern-.08em
    T\kern-.1667em\lower.7ex\hbox{E}\kern-.125emX}}

\begin{document}

\title{Update from Hell: Can Coding Agents Survive Hidden Breakage in Dependency Upgrades?}

\author{
\IEEEauthorblockN{
Zijian Luo\IEEEauthorrefmark{1},
Runzhi He\IEEEauthorrefmark{2},
Pengfei Gao\IEEEauthorrefmark{3},
Yu Kang\IEEEauthorrefmark{3},
Zeqi Lin\IEEEauthorrefmark{3},\\
Minghua Ma\IEEEauthorrefmark{3},
Qingwei Lin\IEEEauthorrefmark{3},
Saravan Rajmohan\IEEEauthorrefmark{3},
and Yongqiang Tian\IEEEauthorrefmark{1}}
\IEEEauthorblockA{\IEEEauthorrefmark{1}Monash University\\
Emails: jack.luo@monash.edu, yongqiang.tian@monash.edu}
\IEEEauthorblockA{\IEEEauthorrefmark{2}Peking University\\
Email: rzhe@pku.edu.cn}
\IEEEauthorblockA{\IEEEauthorrefmark{3}Microsoft\\
Emails: gaopengfei@microsoft.com, kay@microsoft.com, \\
Zeqi.Lin@microsoft.com, minghuama@microsoft.com,\\
qlin@microsoft.com, saravar@microsoft.com}

}

\maketitle

\begin{abstract}
Modern software systems rely heavily on third-party dependencies, but upgrading those dependencies remains a costly maintenance activity. Dependency upgrades do not always preserve the function signatures, type systems, APIs, or runtime semantics assumed by existing code. Consequently, developers often need to perform source code adaptations to accommodate dependency-induced changes. However, such code-level changes are often not explicitly communicated to project maintainers, posing a significant challenge to software reliability.

Meanwhile, coding agents have emerged as a new form of software development tool and are increasingly adopted by developers due to their automation capabilities. In this paper, we introduce \benchmarkname, a benchmark consisting of 203 real-world dependency-upgrade tasks across five package ecosystems spanning five language communities, each involving hidden code-level changes that require source code adaptation. We evaluate ten model harness configurations on \benchmarkname. The best configuration, Copilot CLI with GPT-5.5, solves only 100/203 tasks (49.3\%), with substantial variation across agent harnesses, models, and ecosystems, highlighting an important gap between current agent capabilities and real-world software maintenance needs.
Motivated by incomplete migration and diagnostic errors, we develop
\emph{DepUpgrader}, a lightweight repair skill. On the two best-performing
configurations, Copilot CLI and Codex (both with GPT-5.5), it reports relative
pass rate gains of 15.0\% and 6.1\%, respectively.
\end{abstract}


\section{Introduction}

Third-party dependency reuse is now routine in modern software engineering.
Reuse improves productivity, but creates an ongoing maintenance task:
projects must upgrade dependencies to obtain bug fixes and new
functionality~\cite{bavota2015apache} and to address vulnerabilities in direct
and transitive dependencies~\cite{zerouali2022security}. Yet projects often lag
behind available versions~\cite{kula2018developers,cox2015freshness,decan2018technicallag}.
Compatibility work can complicate adoption, particularly when security fixes
are bundled with functional changes~\cite{pashchenko2020qualitative}.

These changes can cause \emph{upgrade-induced breakage}: a new dependency
version alters the APIs, types, build requirements, or runtime semantics
assumed by downstream code. Such incompatibilities can produce failures that
require source-level adaptation~\cite{venturini2023depended}. Existing test
suites may miss faults introduced through dependency
updates~\cite{hejderup2022trusttests}, leaving incorrect behavior undetected.
The consequences can extend to production. For example, in September 2026,
Spotify reported that an automated dependency upgrade passed existing safety
checks but failed after deployment, affecting end
users~\cite{singer2026spotifyquality}.

Dependency-update bots such as Dependabot~\cite{he2023dependabot} and Renovate
automate version changes and open update pull requests. When an upgrade breaks
downstream code, completing the update also requires adapting that code to the
new dependency contract. Coding agents offer a way to automate this remaining
work: they can inspect a repository, modify source files, and execute tests.
We ask whether coding agents can complete upgrades that require source-level
repair beyond bot-generated version changes.

Answering this question requires repository-level tasks that connect a
dependency upgrade to the breakage it introduces and the repair it requires.
Real upgrade pull requests can mix dependency changes, source adaptations, test
updates, and unrelated maintenance. Building an executable repair task from
such a pull request the refore requires separating these components and
verifying that the upgrade exposes a failure that the source repair resolves.
Held-out tests must then check the upgraded behavior, so that evaluation
measures whether an agent completes the migration across the repository.

We present \benchmarkname, an executable repository-level benchmark that
isolates dependency-upgrade repair. We mine 4,660 candidate dependency-bot
pull requests from real-world GitHub repositories and curate 203 tasks across
five ecosystems, each based on a merged pull request requiring source-level
repair.
We separate dependency changes, developer repairs, and held-out tests, and use
a four-state oracle to validate the link between the upgrade, failure, and
repair.

We evaluate coding agents starting from the pre-upgrade repository and an
upgrade instruction, requiring them to update dependencies and repair the source.
Our study addresses four research questions:
\begin{itemize}
    \item \textbf{RQ1:} How difficult is dependency-upgrade repair for current
    coding agents?
    \item \textbf{RQ2:} How do coding agents perform dependency upgrades?
    \item \textbf{RQ3:} What makes a dependency-upgrade task hard?
    \item \textbf{RQ4:} Why do coding agents fail to complete dependency upgrades,
    and how can their performance be improved?
\end{itemize}

Our results show that the best configuration
solves 100/203 tasks (49.3\%), with performance varying across models, harnesses,
and ecosystems. Failure analysis across all ten configurations shows that agents
often update some affected files but miss others, leaving the migration
incomplete.

To address these failures, we develop \emph{DepUpgrader}, a lightweight skill
that guides agents to consult migration documentation, update affected code
throughout the repository, and test their changes. On the two best-performing
configurations, Copilot CLI and Codex (both with GPT-5.5), the skill reports
relative pass-rate gains of 15.0\% and 6.1\%, respectively.

This paper makes the following contributions:
\begin{itemize}
    \item We introduce \benchmarkname, the first executable, repository-level
    benchmark that isolates dependency-upgrade repair through patch
    decomposition, held-out test patches, and a four-state oracle across five
    package ecosystems.
    \item We develop a construction pipeline that combines dependency-update
    bot mining, path-based hunk classification, automated Docker environment setup,
    leakage controls, and four-state oracle validation to produce validated release
    tasks.
    \item We evaluate mainstream coding agents under containerized,
    reproducible task verification and analyze how agent configurations,
    ecosystems, and failure modes shape performance on dependency-upgrade
    repair, and evaluate whether \emph{DepUpgrader} improves their performance.
\end{itemize}

\section{Motivation}
\label{sec:motivation}

Evaluating dependency-upgrade repair requires observing failures in the
downstream project that uses the upgraded dependency. These failures may be
explicit, such as incompatible APIs, types, or build constraints, or implicit,
where changed runtime semantics cause incorrect behavior even though the code
still compiles~\cite{ochoa2022breakingbad,chen2026breakingchanges}. A benchmark
must therefore provide executable client-level failures, not just a record of
library-level changes.

We first examined existing repository-level agent benchmarks and selected
BeyondSWE~\cite{chen2026beyondswe} because it explicitly includes
dependency-driven migration. BeyondSWE contains tasks drawn from 246 real GitHub
repositories, including a dependency-upgrade slice of 120 tasks. A manual audit
of its upgrade tasks highlighted three task-construction issues.

First, a version bump can appear alongside an unrelated source change or test
update. Selecting a pull request because it changes a dependency can therefore
include repairs driven by ordinary bug fixes, refactoring, or test cleanup.
For example, BeyondSWE labels \emph{groq/openbench\#181} as dependency migration,
but the pull request refactors multiple-choice evaluation and fixes answer
scoring rather than repairing upgrade-induced
breakage.\footnote{\url{https://github.com/groq/openbench/pull/181}.}
Task construction needs to establish which failure and repair are attributable
to the upgrade.

Second, patch-derived problem statements can expose the intended repair before
the agent investigates the upgrade. Even after obvious hints are removed, such
descriptions may reveal replacement APIs, target files, or the direction of the
repair. An upgrade task should specify the requested dependency change while
leaving the agent to discover the new dependency contract and the necessary
source adaptations.

Third, agent repairs need an executable verification target. The dependency
upgrade and held-out tests should expose a failure, and replaying the source
repair should make the same tests pass in the same environment. This
fail-to-pass check gives agent repairs a reproducible target and distinguishes
completed source adaptation from a version change alone.

Together, these observations motivate \benchmarkname's construction
requirements: reproducible upgrade-induced failures, instructions that do not
disclose the developer repair, and held-out tests that validate post-upgrade
behavior. We implement these requirements by separating dependency changes,
source repairs, and test patches, then validating their relationships through
an executable oracle. Section~\ref{sec:method} describes this construction
pipeline.

\section{Dataset Construction}
\label{sec:method}

\subsection{Benchmark Overview}

\benchmarkname\ converts real dependency-update pull requests into executable
dependency-upgrade repair tasks.  A merged pull request is not directly a task:
it may combine dependency-manifest changes, a developer repair, test evolution,
and unrelated maintainer churn.  The goal of the construction pipeline is to
separate these components and retain only tasks where the held-out failure is
causally tied to the dependency upgrade.

Figure~\ref{fig:workflow} summarizes the workflow.  The construction phase
starts from merged pull requests opened by dependency-update bots, performs
patch decomposition to separate manifest changes, repair changes, and a
held-out test patch. Automated Docker setup prepares candidate environments;
the four-state oracle validates the tasks for release. The evaluation phase uses these tasks to test
coding agents: an agent starts from the base repository state and an upgrade
instruction, while the developer repair and held-out test patch remain hidden
until task verification.

\begin{figure*}[t] 
    \centering 
\resizebox{\textwidth}{!}{%
\begin{tikzpicture}[
    x=1cm, y=1cm,
    font=\normalfont\fontsize{8.5}{10}\selectfont,
    text=black, draw=black!75, line width=0.45pt,
    panel/.style={rounded corners=2pt, line width=0.6pt},
    title/.style={anchor=west, font=\normalfont\fontsize{12}{14}\selectfont},
    heading/.style={anchor=west, font=\normalfont\fontsize{9.4}{11}\selectfont},
    body/.style={anchor=west, font=\normalfont\fontsize{8.3}{10}\selectfont},
    metric/.style={anchor=west, font=\normalfont\fontsize{12}{14}\selectfont},
    note/.style={font=\normalfont\fontsize{7}{8.5}\selectfont, align=center},
    panel note/.style={note, text width=4.9cm},
    separator/.style={densely dotted, draw=black!30, line width=0.4pt},
    flow/.style={-{Stealth[length=1.5mm,width=1.2mm]}, draw=black!65, line width=0.55pt},
    step flow/.style={-{Stealth[length=1mm,width=0.8mm]}, draw=black!50, line width=0.4pt},
    pics/pr/.style={code={
        \draw (-0.17,0.24) -- (-0.17,-0.24);
        \draw (0.17,0.16) -- (0.17,-0.12) -- (-0.17,-0.24);
        \foreach \x/\y in {-0.17/0.24,-0.17/-0.24,0.17/0.16}
            \draw[fill=white] (\x,\y) circle (0.065);
    }},
    pics/folder/.style={code={
        \draw[rounded corners=1pt]
            (-0.28,-0.21) -- (-0.28,0.22) -- (-0.03,0.22)
            -- (0.04,0.13) -- (0.28,0.13) -- (0.28,-0.21) -- cycle;
        \draw (-0.28,0.06) -- (0.28,0.06);
    }},
    pics/document/.style={code={
        \draw[rounded corners=0.7pt]
            (-0.21,-0.26) -- (-0.21,0.26) -- (0.08,0.26)
            -- (0.21,0.13) -- (0.21,-0.26) -- cycle;
        \draw (0.08,0.26) -- (0.08,0.13) -- (0.21,0.13);
        \draw (-0.12,0.04) -- (0.12,0.04);
        \draw (-0.12,-0.05) -- (0.12,-0.05);
        \draw (-0.12,-0.14) -- (0.05,-0.14);
    }},
    pics/source-code/.style={code={
        \draw[rounded corners=1.5pt] (-0.23,-0.24) rectangle (0.23,0.24);
        \draw (-0.08,0.09) -- (-0.16,0) -- (-0.08,-0.09);
        \draw (0.08,0.09) -- (0.16,0) -- (0.08,-0.09);
        \draw (0.04,0.12) -- (-0.04,-0.12);
        \draw (-0.13,-0.18) -- (0.13,-0.18);
    }},
    pics/tests/.style={code={
        \draw[rounded corners=1pt] (-0.23,-0.24) rectangle (0.23,0.24);
        \foreach \y in {0.1,-0.1} {
            \draw (-0.16,\y) -- ++(0.04,-0.04) -- ++(0.08,0.09);
            \draw (0.02,\y) -- (0.15,\y);
        }
    }},
    pics/database/.style={code={
        \draw (0,0.2) ellipse [x radius=0.27,y radius=0.08];
        \draw (-0.27,0.2) -- (-0.27,-0.2)
            arc[start angle=180,end angle=360,x radius=0.27,y radius=0.08]
            -- (0.27,0.2);
        \draw (-0.27,0.01)
            arc[start angle=180,end angle=360,x radius=0.27,y radius=0.08];
    }},
    pics/terminal/.style={code={
        \draw[rounded corners=1pt] (-0.28,-0.22) rectangle (0.28,0.22);
        \draw (-0.28,0.12) -- (0.28,0.12);
        \draw (-0.17,0.03) -- (-0.08,-0.04) -- (-0.17,-0.11);
        \draw (-0.01,-0.11) -- (0.14,-0.11);
    }},
    pics/packages/.style={code={
        \draw (-0.12,0.02) rectangle (0.12,0.26);
        \draw (-0.29,-0.27) rectangle (-0.05,-0.03);
        \draw (0.05,-0.27) rectangle (0.29,-0.03);
        \draw (0,0.02) -- (0,-0.01);
        \draw (-0.17,-0.03) -- (-0.17,-0.01) -- (0.17,-0.01) -- (0.17,-0.03);
    }},
    pics/target/.style={code={
        \draw (0,0) circle (0.26);
        \draw (0,0) circle (0.17);
        \draw (0,0) circle (0.075);
        \draw (0,0) -- (0.3,0.3);
        \draw (0.18,0.3) -- (0.3,0.3) -- (0.3,0.18);
    }}
]
\foreach \x in {0,6,12} {
    \draw[panel] (\x,0) rectangle ++(5.5,-7.85);
    \draw (\x,-0.92) -- ++(5.5,0);
}
\pic at (0.48,-0.46) {pr};
\pic at (6.48,-0.46) {database};
\pic at (12.48,-0.46) {target};
\node[title] at (0.95,-0.46) {A. PR Mining};
\node[title] at (6.95,-0.46) {B. Causal Validation};
\node[title] at (12.95,-0.46) {C. Agent Evaluation};

\foreach \y/\label in {
    -1.33/Mine dependency-update PRs,
    -3.50/Decompose merged changes,
    -5.65/Package candidate tasks} {
    \fill[black!75] (0.28,\y-0.11) rectangle ++(0.055,0.22);
    \node[heading] at (0.48,\y) {\label};
}
\foreach \y in {-3.13,-5.28}
    \draw[separator] (0.28,\y) -- (5.22,\y);
\foreach \y in {-3.08,-5.20}
    \draw[step flow] (2.75,\y) -- ++(0,-0.23);
\pic at (0.82,-2.15) {pr};
\node[metric] at (1.40,-1.98) {4,660 candidate PRs};
\node[body] at (1.40,-2.41) {Dependabot / Renovate};
\node[panel note] at (2.75,-2.88) {Bot failed $\rightarrow$ human repair merged};

\pic at (1.00,-4.18) {document};
\pic at (2.75,-4.18) {source-code};
\pic at (4.50,-4.18) {tests};
\node at (1.00,-4.70) {Manifest $m$};
\node at (2.75,-4.70) {Repair $c$};
\node at (4.50,-4.70) {Tests $t$};
\node[panel note] at (2.75,-5.05) {Separate manifest, source, and test changes};

\pic at (0.82,-6.43) {folder};
\node[metric] at (1.40,-6.25) {Pinned base $b$};
\node[body] at (1.40,-6.68) {Environment + oracle $\Omega$};
\node[panel note] at (2.75,-7.24) {Candidate tuple $(b,m,c,t,\Omega)$};

\foreach \y/\label in {
    -1.33/Automated Docker setup,
    -3.50/Apply the four-state oracle,
    -6.11/Release validated benchmark} {
    \fill[black!75] (6.28,\y-0.11) rectangle ++(0.055,0.22);
    \node[heading] at (6.48,\y) {\label};
}
\foreach \y in {-3.13,-5.77}
    \draw[separator] (6.28,\y) -- (11.22,\y);
\foreach \y in {-3.08,-5.70}
    \draw[step flow] (8.75,\y) -- ++(0,-0.23);
\pic at (6.82,-2.15) {terminal};
\node[metric] at (7.40,-1.98) {Docker images};
\node[body] at (7.40,-2.41) {Pinned base + dependencies};
\node[panel note] at (8.75,-2.88) {Automated environment build and setup};
\foreach \y/\state/\result in {
    -4.06/b/Pass,
    -4.44/{b \oplus m \oplus t}/Fail,
    -4.82/{b \oplus m \oplus c \oplus t}/Pass,
    -5.20/{b \oplus c \oplus t}/Fail} {
    \node[anchor=west] at (6.62,\y) {$\Omega(\state)$};
    \node[anchor=east, font=\normalfont\fontsize{8.5}{10}\selectfont]
        at (11.00,\y) {\result};
}
\node[panel note] at (8.75,-5.53) {All four outcomes required in the same round};
\pic at (6.82,-6.87) {database};
\node[metric] at (7.40,-6.65) {\benchmarkname};
\node[body] at (7.40,-7.08) {203 tasks across 5 ecosystems};
\node[panel note] at (8.75,-7.55) {npm/yarn $\cdot$ Maven/Java $\cdot$ Go $\cdot$ Cargo $\cdot$ Python};

\foreach \y/\label in {
    -1.33/Set up agent evaluation,
    -3.50/Run coding agents,
    -5.65/Verify held-out behavior} {
    \fill[black!75] (12.28,\y-0.11) rectangle ++(0.055,0.22);
    \node[heading] at (12.48,\y) {\label};
}
\foreach \y in {-3.13,-5.28}
    \draw[separator] (12.28,\y) -- (17.22,\y);
\foreach \y in {-3.08,-5.20}
    \draw[step flow] (14.75,\y) -- ++(0,-0.23);
\pic at (12.82,-2.16) {terminal};
\node[heading] at (13.40,-1.98) {Base repository $b$};
\node[body] at (13.40,-2.41) {Upgrade instruction};
\draw[dashed, draw=black!40, rounded corners=1pt]
    (12.35,-2.69) rectangle (17.15,-3.01);
\node[note] at (14.75,-2.85) {Hidden: repair $c$, tests $t$, and oracle scripts};

\pic at (13.00,-4.18) {packages};
\pic at (14.75,-4.18) {source-code};
\pic at (16.50,-4.18) {document};
\draw[flow] (13.43,-4.18) -- (14.32,-4.18);
\draw[flow] (15.18,-4.18) -- (16.07,-4.18);
\node at (13.00,-4.70) {Upgrade};
\node at (14.75,-4.70) {Repair};
\node at (16.50,-4.70) {Patch $\hat{c}$};
\node[panel note] at (14.75,-5.05) {Harness--LLM pairs; 60\,min limit};
\node[panel note] at (14.75,-6.10) {Overlay tests $t$; run install / build / tests};
\pic at (12.97,-6.83) {target};
\node[metric] at (13.55,-6.65) {Pass / non-pass};
\node[body] at (13.55,-7.08) {Identical oracle across agents};
\node[panel note] at (14.75,-7.55) {Pass iff the final verifier reward is 1};

\draw[flow] (5.5,-6.43) -- (5.75,-6.43) -- (5.75,-2.36) -- (6,-2.36);
\node[note, rotate=90, fill=white, inner sep=1.5pt] at (5.75,-4.46) {Candidate};
\draw[flow] (11.5,-6.87) -- (11.75,-6.87) -- (11.75,-2.16) -- (12,-2.16);
\node[note, rotate=90, fill=white, inner sep=1.5pt] at (11.75,-4.51) {Task};
\node[note] at (8.75,-8.15)
    {Developer repair $c$, held-out tests $t$, and oracle scripts remain hidden from agents.};
\end{tikzpicture}%
}
    \caption{Overview of \benchmarkname's construction and evaluation workflow.
(A) PR mining and patch decomposition; (B) automated Docker environment setup,
four-state oracle validation, and benchmark release; (C) evaluating agents on
dependency upgrades using held-out tests.}
    \label{fig:workflow} 
    \vspace{-3mm}
\end{figure*}

Each release task packages the information needed for reproducible evaluation:
a pinned repository environment, an upgrade instruction for the agent,
files recording the dependency changes, the developer repair from the original pull
request, a held-out test patch, a task verifier, and metadata such as ecosystem
and resource budget. The input provided to the agent excludes the developer repair and
held-out test patch; those artifacts are used only by the benchmark construction
pipeline and by the task verifier. All upgrade instructions use one template,
without repair summaries or links to the target pull request.

\subsection{PR Mining}
\label{sec:acquisition}

We mine public GitHub pull requests opened by dependency-update bots. The
crawl starts from PRs that are already merged into the main branch and whose
author is a dependency-update bot, such as Dependabot or the Renovate
GitHub App. We then keep only PRs that (i) modify dependency manifests or
lockfiles, (ii) contain non-trivial source or test changes by human
contributors, and (iii) were ultimately integrated into the project's mainline
history.

These filtering criteria ensure, either directly or indirectly, that the retained
candidates satisfy several important properties. First, the PR is related to a
dependency upgrade. Second, the upgrade is non-trivial: if the bot-generated
change were sufficient, the PR could have been merged automatically without
additional intervention. Third, the upgraded dependency introduces observable
differences from the previous version, or changes the downstream behavior enough
to require updates to the source code and tests. Finally, the human effort
invested in resolving the upgrade was accepted by the project maintainers, as
the PR was ultimately merged into the project's mainline history. These
properties make the subsequent filtering and validation steps more tractable.
They are not, however, sufficient to certify a task by themselves: a merged pull
request can still contain unrelated maintainer changes, weak test updates, or
repository bookkeeping. We therefore use mining to identify candidates
and rely on patch decomposition and four-state oracle validation to produce
release tasks.  This crawl identified 4,660 candidate PRs with strong signals
from GitHub, spanning five package ecosystems: npm/yarn, Maven/Java, Go, Cargo,
and Python.

\begin{example}
Figure~\ref{fig:commit-history} shows a representative candidate PR. This PR
(\emph{js-wacz\#41}\footnote{\url{https://github.com/harvard-lil/js-wacz/pull/41}})
was opened by Dependabot to upgrade the project's dependency on \code{glob}
from version 8.1.0 to 10.3.3. The PR contains two commits before it was
eventually merged. The first commit was generated by the bot and updated the
dependency manifest files for the version bump. The second commit was made by a
human contributor and modified the implementation and test files to account for
glob's v10 ESM-style export surface.
\end{example}
\begin{figure*}[t]
\centering
\resizebox{0.95\textwidth}{!}{%
\begin{tikzpicture}[
    font=\scriptsize,
    box/.style={draw=black!65, rounded corners=1pt, align=center,
        minimum height=0.7cm, inner sep=3pt},
    commit/.style={box, text width=3.35cm},
    event/.style={box, text width=2.85cm},
    arrow/.style={-{Stealth[length=2mm]}, thick, draw=black!70},
    brancharrow/.style={-{Stealth[length=2mm]}, thick, dashed, draw=black!70}
]
\node[event] (open) at (0,1.45) {Dependabot opens PR};
\node[commit] (c1) at (3.85,1.45) {PR \#41 commit 1\\857cbde\\Bump glob 8.1.0 $\rightarrow$ 10.3.3\\manifest / lockfile};
\node[commit] (c2) at (8.25,1.45) {PR \#41 commit 2\\12fe776\\Modify source and tests};
\node[event] (review) at (12.25,1.45) {review + CI\\node test\\merged 2023-07-31};

\node[event] (base) at (0,-1.25) {main stream\\167febc\\base before PR \#41\\base branch: main};
\node[commit] (pr39) at (4.65,-1.25) {Another PR lands first\\PR \#39: piscina 3.2.0 $\rightarrow$ 4.0.0};
\node[event] (mainadv) at (8.7,-1.25) {62470d\\main advanced:\\merge PR \#39};
\node[event] (merge41) at (12.25,-1.25) {c176096\\merge PR \#41 into main};

\draw[arrow] (open) -- (c1);
\draw[arrow] (c1) -- (c2);
\draw[arrow] (c2) -- (review);
\draw[arrow] (base) -- (pr39);
\draw[arrow] (pr39) -- (mainadv);
\draw[arrow] (mainadv) -- (merge41);
\draw[brancharrow] (base.north) -- node[left]{branch} (open.south);
\draw[brancharrow] (review.south) -- node[right]{merge} (merge41.north);
\end{tikzpicture}%
}
\caption{Representative commit history for \emph{js-wacz\#41}.}
\label{fig:commit-history}
\end{figure*}

\subsection{Patch Decomposition}

After a PR passes the mining filters, we split its diff into three disjoint
patches.  The \textit{manifest patch} contains dependency declarations and
lockfiles, such as \code{package.json}, \code{go.mod}, \code{go.sum}, or
\code{Cargo.lock}. The
\textit{repair patch} contains the developer repair needed to make the upgraded
dependency work.  The \textit{test patch} contains changed tests, fixtures,
mocks, and expected-output files used by tests.

We first classify hunks by path.  Manifest and lockfile paths are routed to the
manifest patch.  Files under conventional test directories, fixture directories,
or with ecosystem-specific test names are routed to the test patch.
Remaining files are routed to the repair patch.

\begin{example}
The retained PR in Figure~\ref{fig:commit-history} gives a compact example.
For \emph{js-wacz\#41}, the upstream diff is decomposed as follows.

\par\smallskip
\noindent\begin{minipage}{\columnwidth}
\noindent\textbf{Manifest patch.} The dependency declaration and lockfile move
\code{glob} from 8.1.0 to 10.3.3, pinning the new dependency
state in which the v10 ESM entry exposes \code{globSync} as the named
synchronous API.\footnotemark{} The breaking change therefore sits at the import
boundary: code written against the v8 default export object and its
\code{.sync} property must now use the v10 named export.
\begin{lstlisting}[style=diffsnippet]
diff --git a/package.json b/package.json
@@
-    "glob": "^8.1.0",
+    "glob": "^10.3.3",

diff --git a/package-lock.json b/package-lock.json
@@
-        "glob": "^8.1.0",
+        "glob": "^10.3.3",
\end{lstlisting}
\end{minipage}
\footnotetext{See \url{https://github.com/isaacs/node-glob}.}
\par\smallskip

\noindent\begin{minipage}{\columnwidth}
\noindent\textbf{Repair patch.} The source migrates from the old default import
and \code{glob.sync} call to the v10 \code{globSync} API. This patch
records the corresponding source-side adaptation to the new export surface.
\begin{lstlisting}[style=diffsnippet]
diff --git a/index.js b/index.js
@@
-import glob from 'glob'
+import { globSync } from 'glob'
@@
-        for (const file of glob.sync(path)) {
+        for (const file of globSync(path)) {
\end{lstlisting}
\end{minipage}
\par\smallskip

\noindent\begin{minipage}{\columnwidth}
\noindent\textbf{Test patch.} The test patch updates the same upgraded API
surface because the test file also used \code{glob.sync} to compute the
expected fixture count.
\begin{lstlisting}[style=diffsnippet]
diff --git a/index.test.js b/index.test.js
@@
-import glob from 'glob'
+import { globSync } from 'glob'
@@
-  assert.equal(warcCount, glob.sync(FIXTURE_INPUT).length)
+  assert.equal(warcCount, globSync(FIXTURE_INPUT).length)
\end{lstlisting}
\end{minipage}
\par\smallskip
\end{example}
\subsection{Environment Setup, Validation, and Release}
\label{sec:oracle}

\noindent\textbf{Automated environment setup.} We automate candidate environment
setup with Docker, pinning the repository base state and preparing task-specific
install, build, and test commands. The packaged environment supports both
four-state validation and subsequent agent evaluation.

\noindent\textbf{Four-state validation.} Restricting candidates to dependency-update bot pull requests keeps the task
signal clean, but it cannot fully exclude unrelated maintainer context.  For
example, a project may bump its own version after an upgrade; if the test patch
asserts that version, an agent can correctly adapt the upgraded dependency yet
still fail on release bookkeeping incidental to the upgrade.  We therefore
validate every retained candidate with a four-state oracle before release.

\noindent\textbf{Definition 1 (Validation of a dependency-upgrade repair task).}
Let $\mathcal{R}$ be the set of repository states. We write $r \oplus p$ for
the state obtained by applying patch $p$ to state $r$, and let
$\Omega : \mathcal{R}\rightarrow\{\mathsf{pass},\mathsf{fail}\}$ be the task
validation function that runs the task-specific install, build, and test
commands. A candidate tuple $(b,m,c,t,\Omega)$ consists of a base repository
state $b$, manifest patch $m$, repair patch $c$ for the code repair, test patch
$t$, and validation oracle $\Omega$. The candidate passes validation if and only
if all four conditions hold in the same validation round:
\[
\begin{array}{l}
\Omega(b)=\mathsf{pass} \\
\land \Omega(b \oplus m \oplus t)=\mathsf{fail} \\
\land \Omega(b \oplus m \oplus c \oplus t)=\mathsf{pass} \\
\land \Omega(b \oplus c \oplus t)=\mathsf{fail}.
\end{array}
\]
Here, $m$ is restricted to dependency manifests and lockfiles, $c$ is restricted
to the developer repair and must not undo $m$, and $t$ is restricted to tests,
fixtures, mocks, and expected-output files used by tests.  A $\mathsf{fail}$ is
a compilation or build error, a runtime error, or a test failure, rather than
a failure of the evaluation infrastructure. The four checks validate the task under $\Omega$:
$\Omega(b)$ confirms that the base repository state is healthy;
$\Omega(b \oplus m \oplus t)$ shows that the manifest change and test patch
expose an upgrade-induced failure without the repair; $\Omega(b \oplus m
\oplus c \oplus t)$ shows that the repair fixes the upgraded repository under
the same test patch; and $\Omega(b \oplus c \oplus t)$ shows that the repair and
test patch are not sufficient without the dependency upgrade. This last
condition excludes repairs that pass the held-out tests without upgrading.

\noindent\textbf{Benchmark release.} From the 4,660 mined candidate PRs,
\benchmarkname\ releases 203 validated, Docker-packaged tasks (4.4\%).
The release spans five package ecosystems: 68 npm/yarn, 65 Maven/Java, 40 Go,
20 Cargo, and 10 Python tasks.

\section{Empirical Study}
\label{sec:rq}

We evaluate coding-agent configurations on the \benchmarkname\ release using
Harbor,\footnote{\url{https://github.com/harbor-framework/harbor}.} a
containerized evaluation harness for running coding agents on repository-level
tasks. Harbor provisions the pinned task image,
executes the agent under the configured resource budget, applies the held-out test
patch, and computes the final pass/non-pass outcome through the task verifier.  In
this setting, agents
start from the pre-upgrade repository state and must update dependency
manifests or lockfiles, resolve the upgraded dependency state, and repair the
resulting source-level incompatibilities. Each task is counted as a pass when
the final verifier reward is 1 and as a non-pass otherwise. We reserve
\emph{verifier failure} for a non-pass run in which the verifier completes but
the upgraded repository still fails; RQ4 also categorizes non-pass runtime
exceptions.

\subsection{Experiment Setup}

All reported trials run through Harbor in isolated Docker containers, each from a
prebuilt image pinned to the repository base state so every agent sees the same
filesystem, caches, and tooling.  The agent gets the base repository and a
natural-language upgrade instruction, but not the developer repair patch $c$,
hidden test patch $t$, or oracle scripts; the verifier overlays the same hidden test patch $t$ before
grading so all agents face an identical oracle. Harbor blocks the task's
upstream repository and pull-request pages while allowing dependency downloads
and access to dependency documentation. RQ2 examines agents' use of
documentation, release notes, and migration guides.

Trials run on a host with a 128-core CPU and 2\,TB of RAM. The execution protocol
has three phases. First, Harbor starts the
task image and mounts the repository at the base commit.  Second, the agent runs
with a 60\,min time limit and may inspect files, run commands, edit
source/configuration, and perform the dependency upgrade. Third, Harbor invokes
the verifier, which applies
the held-out test patch and runs the task-specific install/build/test command.
Packaged verifier limits are five minutes for 102 tasks and 15--70 minutes for
101 tasks. Missing verifier outcomes count as non-passes. A pass is determined
solely by verifier reward 1, including when the agent previously timed out or
exited with an error. Execution diagnostics are provided in the supplementary
artifact.

We use \emph{harness} for the agent runner or interface without the LLM, and
\emph{configuration} for a harness--LLM pair. We report
GPT-5.5 across Codex, Copilot CLI, OpenCode, and Claude Code; Gemini 3.5 Flash
across Copilot CLI, OpenCode, and Claude Code; and Claude Opus 4.8 across
Copilot CLI, OpenCode, and Claude Code.  Codex is included only on GPT-5.5
because Codex after v0.94.0 dropped the chat-completions wire
API\footnote{\url{https://github.com/openai/codex/pull/10157}.} required by
the non-GPT routes. We selected leading proprietary models and widely used
harnesses available at submission; no open-weight agents were evaluated. The
comparisons concern complete configurations, including tools, reasoning
settings, execution policies, and resource budgets.

The main evaluation reports one trial for each of the 203
tasks under each of the ten configurations (2,030 runs). Performance and
trajectory-behavior statistics use these same trials, matched by configuration,
task, and trial ID. Task-level analyses use the same 203-task release.
The RQ4 failure-category analysis covers all 1,262 non-pass trajectories
across the ten configurations.

\begin{samepage}
We organize the study around four research questions:
\begin{itemize}
    \item \textbf{RQ1:} How difficult is dependency-upgrade repair for current
    coding agents?
    \item \textbf{RQ2:} How do coding agents perform dependency upgrades?
    \item \textbf{RQ3:} What makes a dependency-upgrade task hard?
    \item \textbf{RQ4:} Why do coding agents fail to complete dependency upgrades,
    and how can their performance be improved?
\end{itemize}
\end{samepage}

\subsection{RQ1: How Difficult Is Dependency-Upgrade Repair for Current Coding
Agents?}

\begin{table}[t]
\centering
\caption{\normalfont Success rates by model and agent harness. Dashes denote unsupported combinations.}
\label{tab:agent-comparison}
\footnotesize
\setlength{\tabcolsep}{2pt}
\begin{tabular}{lrrrr}
\toprule
\textbf{Model} & \textbf{Claude Code} & \textbf{Codex} &
\textbf{Copilot CLI} & \textbf{OpenCode} \\
\midrule
GPT-5.5 & 70 (34.5\%) & 99 (48.8\%) & \textbf{100 (49.3\%)} & 81 (39.9\%) \\
Claude Opus 4.8 & 84 (41.4\%) & -- & 88 (43.3\%) & 79 (38.9\%) \\
Gemini 3.5 Flash & 57 (28.1\%) & -- & 57 (28.1\%) & 53 (26.1\%) \\
\bottomrule
\end{tabular}
\end{table}

The main result is that dependency-upgrade repair remains difficult even for the
strongest configurations (Table~\ref{tab:agent-comparison}). The best
configuration, Copilot CLI with GPT-5.5, solves 100/203 tasks (49.3\%); Codex
with GPT-5.5 solves 99/203, OpenCode solves 81/203, and Claude Code solves
70/203. The best configuration therefore still fails to solve a
majority of the release.  Claude Opus 4.8 is competitive but also far from
saturating the benchmark: Copilot CLI, OpenCode, and Claude Code solve 88, 79,
and 84 tasks respectively. Gemini 3.5 Flash has fewer passes in every matched
harness, with 53, 57, and 57 passes for OpenCode, Copilot CLI, and Claude Code.
The best rate has a Wilson 95\% confidence interval of 42.5--56.1\%; a
10,000-replicate repository-cluster bootstrap gives 41.9--56.4\%, accounting for
tasks drawn from the same repository. We use exact McNemar tests for all 45
paired configuration comparisons with Holm correction. With GPT-5.5, Copilot
CLI versus Codex gives $p=1.0$, so the one-task ranking difference does not
establish a performance advantage. Copilot CLI exceeds Claude Code by
14.8 percentage points, with a bootstrap 95\% interval of 8.3--21.0 points.

Holding GPT-5.5 fixed, the observed harness spread is 30 tasks (14.8 percentage
points); it is four tasks (2.0 points) for Gemini and nine tasks (4.4 points) for
Claude Opus 4.8.
Across all ten configurations, 65/203 tasks have zero observed passes.

\begin{table}[t]
\centering
\caption{\normalfont Pass counts by ecosystem for each configuration.
Ecosystem totals are npm/yarn 68, Maven/Java 65, Go 40, Cargo 20, and
Python 10.}
\label{tab:rq1-ecosystem}
\scriptsize
\setlength{\tabcolsep}{2pt}
\begin{tabular}{lrrrrr}
\toprule
\textbf{Configuration} & \textbf{npm} & \textbf{Maven} &
\textbf{Go} & \textbf{Cargo} & \textbf{Python} \\
\midrule
Copilot CLI + GPT-5.5 & 27 & 38 & 20 & 9 & 6 \\
Codex + GPT-5.5 & 26 & 40 & 20 & 8 & 5 \\
OpenCode + GPT-5.5 & 23 & 26 & 18 & 8 & 6 \\
Claude Code + GPT-5.5 & 24 & 22 & 13 & 6 & 5 \\
\midrule
Copilot CLI + Claude Opus 4.8 & 21 & 34 & 20 & 8 & 5 \\
OpenCode + Claude Opus 4.8 & 18 & 29 & 20 & 7 & 5 \\
Claude Code + Claude Opus 4.8 & 21 & 28 & 23 & 7 & 5 \\
\midrule
OpenCode + Gemini 3.5 Flash & 11 & 20 & 12 & 7 & 3 \\
Copilot CLI + Gemini 3.5 Flash & 15 & 23 & 10 & 6 & 3 \\
Claude Code + Gemini 3.5 Flash & 13 & 23 & 11 & 6 & 4 \\
\bottomrule
\end{tabular}
\end{table}

Observed success rates vary by ecosystem (Table~\ref{tab:rq1-ecosystem}).
With GPT-5.5, Copilot CLI solves 38/65 Maven/Java tasks, Codex 40/65,
OpenCode 26/65, and Claude Code 22/65. Cargo passes range from 6/20 to 9/20,
while each configuration solves fewer than half of the 68 npm/yarn tasks.
Python has only ten tasks, so its rates are particularly sensitive to individual
outcomes.

Model choice also interacts with the observed ecosystem results. For Copilot
CLI, GPT-5.5 versus Gemini 3.5 Flash gives 38 versus 23 Maven/Java passes,
20 versus 10 Go passes, 27 versus 15 npm/yarn passes, 6 versus 3 Python passes,
and 9 versus 6 Cargo passes. Claude Opus 4.8 falls between these Copilot CLI
rows overall, with 88 passes, including 34 in Maven/Java.

\rqanswer{Dependency-upgrade repair remains difficult for current coding
agents: the highest observed success rate is 49.3\%, and 65 tasks have no pass
across any of the evaluated configurations.}

\subsection{RQ2: How Do Coding Agents Perform Dependency Upgrades?}

We analyze how agents investigate dependency upgrades and validate their edits
using automated trajectory annotations. The four indicators record dependency
documentation lookup, local dependency API/source inspection, test/build
execution, and explicit breaking-change hypotheses. All 2,030 behavior
annotations align with the final reported trials. The indicators have separate
Cohen's $\kappa$ values ranging from 0.613 to 0.864 on 748--749 matched annotation
pairs per indicator. Table~\ref{tab:rq2-trajectory} reports counts of these
non-exclusive behaviors.

\begin{table}[t]
\centering
\caption{\normalfont Counts of discovery behaviors observed in the final
trajectories (203 per configuration). Columns are non-exclusive. Docs counts
dependency documentation, release notes, or migration evidence, excluding
searches for fixes to the task repository; API means local dependency API/source inspection; BC
hyp. means an explicit breaking-change hypothesis.}
\label{tab:rq2-trajectory}
\footnotesize
\begin{tabular*}{\columnwidth}{@{\extracolsep{\fill}}lrrrr@{}}
\toprule
\textbf{Configuration} & \textbf{Docs} & \textbf{API} &
\textbf{Tests} & \textbf{BC hyp.} \\
\midrule
Copilot CLI + GPT-5.5 & 18 & 151 & 169 & 181 \\
Codex + GPT-5.5 & 97 & 172 & 189 & 191 \\
OpenCode + GPT-5.5 & 72 & 155 & 161 & 163 \\
Claude Code + GPT-5.5 & 47 & 178 & 193 & 169 \\
\midrule
OpenCode + Gemini 3.5 Flash & 59 & 87 & 115 & 76 \\
Copilot CLI + Gemini 3.5 Flash & 48 & 77 & 130 & 77 \\
Claude Code + Gemini 3.5 Flash & 120 & 135 & 197 & 152 \\
\midrule
Copilot CLI + Claude Opus 4.8 & 21 & 167 & 168 & 179 \\
OpenCode + Claude Opus 4.8 & 53 & 161 & 169 & 166 \\
Claude Code + Claude Opus 4.8 & 79 & 180 & 186 & 187 \\
\bottomrule
\end{tabular*}
\end{table}

Two descriptive patterns are clear. First, external documentation lookup varies
sharply by configuration and is not the dominant signal overall. For GPT-5.5,
documentation evidence ranges from 18 to 97 trajectories across configurations,
while local dependency API/source inspection appears in 151 to 178 trajectories
and test execution appears in 161 to 193 trajectories. Claude Opus 4.8
shows a similarly broad spread in documentation behavior (21 to 79 trajectories)
while retaining high counts for local API/source inspection and test/build
feedback. Second, explicit breaking-change hypotheses appear frequently once
agents start reasoning about failures.

\rqanswer{Agents tackle upgrades mainly through local dependency API
inspection and test/build feedback; documentation lookup is less common.}

\subsection{RQ3: What Makes a Dependency-Upgrade Task Hard?}

The results suggest that task difficulty is not explained by a single factor.
We examine ecosystem tools, the behavior checked by held-out
tests, and the need to update code throughout a repository. For example,
10/20 Cargo tasks and 34/68 npm/yarn tasks have no pass among the four
GPT-5.5 configurations; the corresponding counts are 10/20 and 45/68 for
Gemini, and 11/20 and 43/68 for Claude Opus 4.8. These observations motivate
closer inspection of tooling and migration requirements.

We next examine what the held-out tests check. To characterize the upgrade
requirements facing agents, we audit two aspects of the test patches
(Table~\ref{tab:rq3-test-quality}): whether they directly assert or correct
dependency-induced behavior, and whether they distinguish a source-level repair
from making no changes or simply deleting tests.

\begin{table}[t]
\centering
\caption{\normalfont Held-out test behavior labels for the current 203-task release.
Tasks may receive multiple labels in the upper block; the lower block assigns
one primary label per task.}
\label{tab:rq3-test-quality}
\footnotesize
\setlength{\tabcolsep}{4pt}
\begin{tabular}{lrr}
\toprule
\textbf{Held-out test signal} & \textbf{Tasks} & \textbf{Share} \\
\midrule
Corrects existing behavior & 142 & 70.0\% \\
Adds new tests or coverage & 60 & 29.6\% \\
Deletes/removes tests & 39 & 19.2\% \\
Fixture/setup only & 47 & 23.2\% \\
\midrule
Primary: corrects only & 87 & 42.9\% \\
Primary: mixed changes & 55 & 27.1\% \\
Primary: adds only & 14 & 6.9\% \\
Primary: fixture/setup only & 47 & 23.2\% \\
\bottomrule
\end{tabular}
\end{table}

Table~\ref{tab:rq3-test-quality} summarizes these labels. Most tasks
have direct behavioral signal: 142/203 modify existing expectations, mocks,
fixtures, or snapshots to match the upgraded dependency, and 60/203 add new
tests or test cases. The upper-block labels are non-exclusive: a patch can
both remove obsolete tests and add replacement coverage.
The lower block assigns each task exactly one primary label. In this grouping,
156/203 tasks (76.8\%) either
correct existing behavior, add direct coverage, or combine both with smaller
deletions. These are the strongest tasks because the hidden test patch forces
the agent to reproduce observable behavior of the upgraded dependency rather
than merely keep the old program compiling.

Tasks whose test patches combine several types of behavioral changes have
the highest proportion of unsolved tasks: 29/55 are solved by none of the four
GPT-5.5 configurations. Tasks whose tests only correct existing behavior are
also difficult, with 36/87 remaining unsolved. The counts are lower for tasks
that only add tests or only change fixtures or test setup, at 4/14 and 8/47
respectively. One possible explanation is that combining different test changes
or extensively revising existing tests often requires updating mocks, expected
values, wrappers, generated fixtures, and source call sites together.
Test patch size shows a similar association: under GPT-5.5,
tasks solved by no GPT-5.5 configuration have an average of 76.3 changed
hidden-test lines across 4.1 files, versus 36.7 lines across 2.2 files when all
four pass. This association is consistent with broader changes to behavior or
fixtures.

Some tasks require adapting an upgraded API surface directly. In
\emph{29k\#3418}, the OpenAI SDK migration from v3 to v4 changes the client
construction and chat-completion call path; the hidden Jest test rewrites its
mock around the v4 client shape, so a repair that only changes imports cannot pass. In
\emph{29k\#5244}, the PostHog React Native upgrade changes the adapter from
asynchronous initialization to constructor-based usage, and the test checks both
construction and downstream event methods.

Other tasks preserve the same high-level feature but change the expected
behavior. \emph{ArcadeDB\#2808} updates a Gremlin overflow expectation after a
TinkerPop upgrade, so the repair must reflect a changed engine behavior rather
than merely compile. An example involving generated output appears in
\emph{create-typescript-app\#1489}: the hidden test updates the generated package
metadata from pnpm 8 to pnpm 9, forcing the source generator to follow the
package-manager upgrade.

A third group exposes the upgrade through compilation, generated artifacts, or
reactivated regressions. \emph{AxonFramework\#1654} moves validation code from the
Javax namespace to Jakarta, so tests compiled against the new namespace fail
unless production imports and wrappers migrate consistently. In \emph{actix\#554},
new trybuild cases exercise the syn 2 parser contract for proc macros; fixing a
single compile error is not enough if another derive path remains on the old
parser model. \emph{Graylog-server\#21446} and \emph{Spoon\#6087} illustrate still other
forms: generated protobuf sources must stay synchronized with the upgraded
runtime, and a re-enabled JDT regression test forces the integration logic to
match the new dependency behavior. These examples show that a dependency
upgrade witness may be a runtime assertion, a snapshot, a compile path, a
generated artifact, or a focused regression test.

Patches that only change fixtures or test setup expose upgrade effects through serialization,
generated fixtures, package layout, or test bootstrap behavior.
Compilation and fixture tests can expose upgrade-induced failures without new
runtime assertions. We evaluate test removals together with replacement coverage
and the observed failure. The four-state oracle (Section~\ref{sec:oracle})
checks that both the dependency upgrade and the source repair are necessary
under the held-out tests.

During curation, we consider whether all four validation conditions hold,
whether the patch is restricted to tests and supporting files, whether it adds
or updates concrete assertions, and whether the tests exercise the upgraded
dependency surface. These checks focus validation on the dependency contract.

We also check sensitivity to held-out test quality. Copilot CLI with GPT-5.5
has the highest observed rate in each reported subset: 66/156 (42.3\%) on
tasks whose tests check dependency behavior directly, 60/138 (43.5\%) with
unchanged test patches, and 47/103 (45.6\%) after also excluding patches labeled
as removing tests. Thus the
overall leading configuration remains the same under these filters, although
the subsets represent different task populations.

Finally, repairs may need to update several parts of a repository. The two
case studies in RQ4 illustrate this requirement. The GitLab client case requires
propagating type changes from the dependency API through local interfaces,
helper functions, and returned values. The Gremlin case requires connecting a change
in the dependency's numeric behavior to changes in the local comparator and
the user-visible overflow behavior.
In both cases, a plausible local edit is insufficient.  The agent must carry the
upgraded API or semantic requirement through every repository layer exercised by
the hidden oracle.

\rqanswer{Upgrades that affect multiple aspects of behavior place greater
demands on repair. Agents must propagate API and semantic changes consistently
across affected components, rather than fix isolated call sites. 52.7\% of tasks whose test patches combine behavioral
changes have no pass across the four GPT-5.5 configurations, compared with
17.0\% of tasks that only change fixtures or test setup.}

\subsection{RQ4: Why Do Coding Agents Fail to Complete Dependency Upgrades,
and How Can Their Performance Be Improved?}
\label{sec:rq4}

When agents produce a non-pass outcome on \benchmarkname, we examine whether
they identify the relevant dependency change, apply the appropriate API or
semantic adaptation, and update all affected parts of the repository. Across
non-pass runs with a completed verifier outcome, a common challenge is
completing a consistent source repair throughout the repository.

The failure-category analysis covers all 1,262 final non-pass trajectories
across the ten configurations. For each configuration, the number of failures
is 203 minus its passes in Table~\ref{tab:agent-comparison}.
Grok 4.7 and GPT-6.1 Sol independently annotate each trajectory using the
execution trace, final patch, oracle patch, and verifier evidence. They may
assign multiple labels from the eight categories below to the same trajectory.
Agreement between the two annotators before adjudication is Cohen's
$\kappa = 0.5240$. GPT-6 Astra adjudicates disagreements.
Table~\ref{tab:rq4-failure-categories} reports the resulting counts.

\emph{Incomplete migration} covers attempts to upgrade the relevant dependency
with identifiable omissions or inconsistencies in source code, dependency
state, wrappers, or generated artifacts. It includes compilation, type, and
namespace inconsistencies arising from these migration omissions or
inconsistent changes. The categories describe observed failures and may
co-occur within a trajectory; they are not mutually exclusive explanations.

\emph{Wrong root cause} denotes misidentifying the source or nature of the
breakage and basing the repair on an unrelated diagnosis or wrong compatibility
surface---for example, treating an API removal as a network-configuration issue.
It requires positive recorded evidence linking that diagnosis to the repair;
a non-pass verdict, missed edit, or instruction/verifier mismatch alone is
insufficient. Omitted migration steps are labeled as incomplete migration.
\emph{Wrong semantic output} describes an implemented adaptation that executes
but produces incorrect post-upgrade behavior. A trajectory can receive both
labels if it contains an omitted migration step and an incorrect behavioral
change. An unsuccessful implementation alone is not a wrong diagnosis.

The other categories distinguish observable execution failures and strategies:
\emph{runtime exception} records supported runner, authentication, service, or
environment failures preventing meaningful evaluation; \emph{runtime timeout}
records a run that exceeds its execution time limit;
\emph{no upgrade attempted} requires evidence of no effective upgrade or the
retained old dependency, not merely an unavailable diff;
\emph{install/build failure} concerns dependency resolution or non-source
build/bootstrap gates, not residual consumer compiler errors; and
\emph{test tampering} requires decisive test weakening or bypass, not incidental
test edits. Exception markers alone do not establish these explanations.

Incomplete migration is the most frequent label, appearing in 460/1,262
failure trajectories (36.5\%), followed by wrong semantic output in 397
(31.5\%) and wrong root cause in 361 (28.6\%). The trajectories receive
1,919 labels in total. These percentages use the 1,262 failure trajectories
as the denominator and need not sum to 100\%.

\begin{table*}[t]
\centering
\caption{\normalfont Failure labels for all 1,262 non-pass trajectories across
the ten configurations. $N$ is the number of failure trajectories; each
remaining cell counts trajectories with that label. Multiple labels may apply
to one trajectory, so label counts in a row may sum to more than $N$.}
\label{tab:rq4-failure-categories}
\scriptsize
\setlength{\tabcolsep}{3pt}
\begin{tabular}{@{}llrrrrrrrrr@{}}
\toprule
\textbf{Model} & \textbf{Harness} & \textbf{$N$} &
\shortstack{\textbf{Incomplete}\\\textbf{migration}} &
\shortstack{\textbf{Wrong}\\\textbf{semantic}\\\textbf{output}} &
\shortstack{\textbf{Wrong}\\\textbf{root cause}} &
\shortstack{\textbf{Install/build}\\\textbf{failure}} &
\shortstack{\textbf{Runtime}\\\textbf{timeout}} &
\shortstack{\textbf{No upgrade}\\\textbf{attempted}} &
\shortstack{\textbf{Runtime}\\\textbf{exception}} &
\shortstack{\textbf{Test}\\\textbf{tampering}} \\
\midrule
\multirow{3}{*}{Claude Opus 4.8} & Claude Code & 119 & 46 & 43 & 27 & 29 & 13 & 1 & 4 & 8 \\
 & Copilot CLI & 115 & 37 & 57 & 27 & 19 & 15 & 7 & 6 & 11 \\
 & OpenCode & 124 & 49 & 49 & 28 & 9 & 28 & 19 & 1 & 6 \\
\midrule
\multirow{3}{*}{Gemini 3.5 Flash} & Claude Code & 146 & 49 & 42 & 47 & 33 & 22 & 7 & 30 & 12 \\
 & Copilot CLI & 146 & 78 & 22 & 64 & 14 & 8 & 20 & 8 & 9 \\
 & OpenCode & 150 & 40 & 24 & 41 & 39 & 28 & 70 & 7 & 7 \\
\midrule
\multirow{4}{*}{GPT-5.5} & Claude Code & 133 & 39 & 40 & 39 & 38 & 21 & 2 & 6 & 4 \\
 & Codex & 104 & 37 & 41 & 31 & 11 & 6 & 3 & 7 & 6 \\
 & Copilot CLI & 103 & 43 & 44 & 25 & 9 & 10 & 1 & 4 & 4 \\
 & OpenCode & 122 & 42 & 35 & 32 & 14 & 36 & 24 & 1 & 4 \\
\midrule
\textbf{Total} & & \textbf{1262} & \textbf{460} & \textbf{397} & \textbf{361} &
\textbf{215} & \textbf{187} & \textbf{154} & \textbf{74} & \textbf{71} \\
\bottomrule
\end{tabular}
\end{table*}

The label distribution also varies across configurations. For example,
incomplete migration appears in 78/146 failures (53.4\%) for Copilot CLI with
Gemini 3.5 Flash, while no upgrade is attempted in 70/150 failures (46.7\%)
for OpenCode with the same model.

For each LLM, we group tasks by whether none, some, or all of its evaluated
configurations pass. Specifically, 77/203 tasks for GPT-5.5, 107/203 for Gemini,
and 94/203 for Claude Opus 4.8 are not solved by any evaluated configuration
using that model. Tasks solved by some configurations
but not others help identify differences in how agents perform repairs. In
these cases, an agent may identify the relevant dependency change but miss
affected wrappers, types, helpers, fixtures, or call sites. The case studies
below illustrate such incomplete repairs and errors in post-upgrade behavior.

\begin{table}[t]
\centering
\caption{\normalfont Tasks solved by none, some, or all evaluated configurations,
by LLM and ecosystem. Partial pass means at least one but not all evaluated
configurations using the same LLM pass.}
\label{tab:rq4-overlap-eco}
\scriptsize
\setlength{\tabcolsep}{3pt}
\begin{tabular}{llrrrrr}
\toprule
\textbf{LLM} & \textbf{Ecosystem} & \textbf{Harnesses} & \textbf{N} &
\textbf{0 pass} & \textbf{Partial pass} & \textbf{All pass} \\
\midrule
\multirow{5}{*}{GPT-5.5} & npm/yarn & 4 & 68 & 34 & 17 & 17 \\
 & Maven/Java & 4 & 65 & 16 & 37 & 12 \\
 & Go & 4 & 40 & 13 & 19 & 8 \\
 & Cargo & 4 & 20 & 10 & 5 & 5 \\
 & Python & 4 & 10 & 4 & 2 & 4 \\
\midrule
\multirow{5}{*}{Gemini 3.5 Flash} & npm/yarn & 3 & 68 & 45 & 21 & 2 \\
 & Maven/Java & 3 & 65 & 25 & 36 & 4 \\
 & Go & 3 & 40 & 23 & 12 & 5 \\
 & Cargo & 3 & 20 & 10 & 7 & 3 \\
 & Python & 3 & 10 & 4 & 4 & 2 \\
\midrule
\multirow{5}{*}{Claude Opus 4.8} & npm/yarn & 3 & 68 & 43 & 11 & 14 \\
 & Maven/Java & 3 & 65 & 20 & 32 & 13 \\
 & Go & 3 & 40 & 15 & 9 & 16 \\
 & Cargo & 3 & 20 & 11 & 3 & 6 \\
 & Python & 3 & 10 & 5 & 0 & 5 \\
\bottomrule
\end{tabular}
\end{table}

The number of tasks that no configuration solves varies by ecosystem
(Table~\ref{tab:rq4-overlap-eco}). npm/yarn contributes the most such tasks:
under GPT-5.5, 34/68 are unsolved by all four
configurations, under Gemini this is 45/68 among the three included configurations,
and under Claude Opus 4.8 it is 43/68. Cargo has 10/20 tasks that no configuration
solves under GPT-5.5 and Gemini and 11/20 under Claude Opus 4.8. Many Maven/Java
tasks are solved by some configurations but not others: 37/65 under GPT-5.5 and
32/65 under Claude Opus 4.8.

The held-out tests also vary by kind, so agents must repair different failure
mechanisms rather than apply a single migration recipe. For tests of changed
APIs, a common failure is \emph{incomplete migration}: the agent replaces
one removed symbol but misses a wrapper, mock, factory, return object, or
secondary call site.  The OpenAI and PostHog examples in RQ3 require the repair
to migrate both construction and downstream method calls; a patch that changes
only the import or only the constructor still fails the hidden mock assertions.

For tests of output behavior, agents often make a syntactically valid migration but
do not re-derive the changed behavior.  A Gremlin repair can compile yet fail the
new overflow expectation; a repair to a package generator can update a dependency
declaration but still emit the old package-manager snapshot.  These
failures are not simply caused by missing imports. They show that the agent did not connect
the dependency version change to the observable output encoded by the test.

Compilation and namespace failures can reflect an incomplete
migration: agents fix the first visible compiler error but leave the
migration inconsistent across the repository. For the Javax-to-Jakarta
validation migration, for example, tests
compile against the new namespace while production code may still refer to the
old one.  For syn macro migrations, a local parsing fix can
leave another derive path or compile-fail expectation broken.  Fixture/setup
tests add another source of failure: generated sources, lockfiles, snapshots, and
test bootstrap files must be updated consistently with production code.
Compilation, fixture, and runtime-assertion tests are witness types, not
additional peer failure categories.

\noindent\textbf{Case Study 1: signature change propagation.}
One representative propagation case is
\emph{renoglaab\#30}, which upgrades the Go GitLab client
from v0.161.1 to v1.2.0.  This is a
clear breaking change: GitLab merge-request and pipeline identifiers exposed by
the client move from int to int64.  The repair therefore has to
propagate the widened ID type through the repository's GitLab abstraction and
merge-request filtering logic.  The official 1.0 migration guide instructs
clients to migrate references accordingly,\footnote{GitLab client-go Version 1.0
Migration Guide, ``Migrate int to int64'':
\url{https://gitlab.com/gitlab-org/api/client-go/-/blob/v1.2.0/docs/release-1.0-migration.md\#migrate-int-to-int64}.}
and the module source confirms that the concrete surfaces used here---pipeline
IDs, merge-request IIDs, and GetPipeline's pipeline parameter---changed from int
to int64 between v0.161.1 and v1.2.0.

\par\addvspace{\medskipamount}
\noindent\hspace*{-1.1pt}%
\setlength{\fboxsep}{5pt}%
\fbox{\begin{minipage}{\dimexpr\linewidth-2\fboxsep-2\fboxrule\relax}
\normalsize
\textbf{client-go migration guide excerpt.}
``All structs and functions in `client-go` have been updated to use `int64`
types where they previously used `int`. This ensures there is no ambiguity about
the fact that GitLab can use large integers to all integer values, and prevents
some cross-platform issues. When referencing `int64` returns from `client-go`,
either update the referencing code to use `int64`, or cast it to an `int`
instead.''
\end{minipage}}
\par\addvspace{\medskipamount}

\noindent\begin{minipage}{\linewidth}
\begin{lstlisting}[style=diffsnippet]
// internal/gitlab/client.go
- GetPipeline(repo string, pipelineID int)
+ GetPipeline(repo string, pipelineID int64)

// internal/mergerequests/list.go
- func listProjectMergeRequests(...) []int
+ func listProjectMergeRequests(...) []int64
\end{lstlisting}
\end{minipage}

In the selected Codex and Copilot CLI trials with GPT-5.5, both agents partially
migrated the client wrapper to int64 but left the merge-request helper
converting identifiers back to int and returning an int slice. The held-out
tests require the new type throughout, exposing the remaining mismatch:

\begin{lstlisting}[style=diffsnippet]
expected: []int64([]int64{1})
actual  : []int([]int{1})
\end{lstlisting}

This case illustrates the benchmark's target failure mode.  The task is not a
simple manifest update: the agent must trace a type change in a dependency API
through interfaces, wrappers, helper functions, and returned values. Both
selected repairs failed to preserve the int64 contract end-to-end, despite
updating relevant parts of the repository.

\noindent\textbf{Case Study 2: semantic behavior change in Gremlin.}
\emph{ArcadeDB\#2808} illustrates a different kind of breaking change: the public
surface is not the main issue; the dependency's runtime semantics changed.  The
task upgrades the Gremlin/TinkerPop dependency from 3.7.4 to 3.8.0.  Under the
new engine, adding one to the maximum long value is no longer treated like the
old behavior expected by ArcadeDB's test.  The TinkerPop 3.8.0 upgrade guide
lists this as a breaking change in numeric behavior: integer operations now promote
byte to short to int to long and then throw an overflow exception; its example
specifically shows a 3.8.0 traversal over Long.MAX\_VALUE plus one throwing
``java.lang.ArithmeticException: long overflow.''\footnote{Apache TinkerPop
3.8.0 Upgrade Guide, ``Auto-promotion of Numbers'':
\url{https://tinkerpop.apache.org/docs/3.8.0/upgrade/\#auto-promotion-of-numbers}.}
The hidden test mirrors that upstream guidance by expecting an arithmetic
overflow exception for the long-valued expression, while explicitly using a
big-integer value still succeeds.  In other words, the oracle is not asking the
agent to rename an API.  It asks whether the repository now respects the
upgraded engine's numeric semantics.

\par\addvspace{\medskipamount}
\noindent\hspace*{2pt}%
\setlength{\fboxsep}{5pt}%
\fbox{\begin{minipage}{\dimexpr\linewidth-4pt-2\fboxsep-2\fboxrule\relax}
\normalsize
\textbf{TinkerPop 3.8.0 Upgrade Guide excerpt.}
``Now, any mathematical operations such as `Add`, `Sub`, `Mul`, and `Div` will
now automatically promote to the next numeric type if an overflow is detected.
For integers, the promotion sequence is: byte → short → int → long → overflow
exception. For floating-point numbers, the sequence is: float → double →
infinity. The following example showcases the change in overflow behavior between
3.7.3 and 3.8.0.''  ``// 3.8.0 gremlin>
g.inject([Long.MAX\_VALUE, 1l]).sum(local) // throws
java.lang.ArithmeticException: long overflow.''
\end{minipage}}
\par\addvspace{\medskipamount}

\noindent\begin{minipage}{\linewidth}
\begin{lstlisting}[style=diffsnippet]
// Simplified from GremlinTest.longOverflow()
- result = gremlin("Long.MAX_VALUE + 1")
- assertThat(result).isEqualTo(Long.MAX_VALUE + 1)
+ assertThatThrownBy(() -> gremlin("Long.MAX_VALUE + 1"))
+     .isInstanceOf(ArithmeticException.class)
+     .hasMessageContaining("long overflow");

// Explicit BigInteger summation remains valid.
  gremlin("BigInteger(Long.MAX_VALUE) + 1")
\end{lstlisting}
\end{minipage}

The source repair is also subtle.  ArcadeDB carries a Gremlin value comparator
that must remain compatible with TinkerPop internals.  The upstream repair stops
depending on a removed exception for type errors defined by Gremlin, switches to a
standard exception path, and exposes a comparability helper that the upgraded
comparison logic can access.  These changes are small in file count, but they
connect two aspects of the repair: compatibility with the library's comparison
logic, and the user-visible overflow behavior checked by the hidden test.

The selected Codex and Copilot CLI trials with GPT-5.5 illustrate the semantic
error directly. Both added or modified a local NumberHelper override to preserve
the old integer-overflow behavior. The hidden overflow test instead expects the
upgraded engine to throw, so these repairs fail despite attempting to restore
compatibility. The GitLab repairs above omit required type changes, whereas
these Gremlin repairs incorrectly preserve the old overflow behavior.

\textbf{Preliminary skill-guided mitigation.} The GitLab and Gremlin cases above
highlight two complementary needs: propagating changes across affected code
and checking behavior against the upgraded contract rather than preserving
obsolete semantics. We translate these needs into \emph{DepUpgrader}, a
lightweight, task-independent skill that combines version-matched migration
evidence, affected-site searches, and final validation. It provides procedural
guidance, not task-specific repair patches or held-out test information.

\par\addvspace{\medskipamount}
\noindent\begingroup
\setlength{\fboxsep}{5pt}%
\fbox{\begin{minipage}{\dimexpr\linewidth-2\fboxsep-2\fboxrule\relax}
\small
\textbf{DepUpgrader}
\par\smallskip\noindent
\textbf{Confirm the upgrade}: Apply or confirm the target version in the
manifest and lockfile.
\textbf{Extract migration rules}: Read relevant, version-matched upstream
docs or installed API/types. Record
\texttt{old API -> new API | required edit | evidence}. Do not guess replacements.
\textbf{Find affected code}: Search imports, calls, and old symbols. Check
aliases and local wrappers when needed.
\textbf{Make minimal edits}: Update affected imports, calls, types, and
wrappers consistently. Adapt usage flow for behavior or lifecycle changes.
Leave compatible code unchanged.
\textbf{Validate}: Confirm the resolved version, run focused build/type/test
checks after the final edit, search for remaining old usages, and inspect the diff.
\end{minipage}}
\endgroup
\par\addvspace{\medskipamount}

The DepUpgrader study with Copilot CLI and GPT-5.5 reports a pass rate of
56.7\% (115/203 tasks; Table~\ref{tab:skill-pilot}), compared with
49.3\% (100/203 tasks) in the original baseline evaluation
(Table~\ref{tab:agent-comparison}). Aggregation details are provided in the
accompanying analysis documentation.

Also with Codex and GPT-5.5, DepUpgrader reports a pass rate of 51.7\%
(105/203 tasks), compared with 48.8\% (99/203 tasks) in the original baseline,
a difference of six passes.

\begin{table}[t]
\centering
\caption{\normalfont Pass rate in the DepUpgrader study with GPT-5.5.}
\label{tab:skill-pilot}
\footnotesize
\setlength{\tabcolsep}{4pt}
\begin{tabular}{lrrr}
\toprule
\textbf{Harness} & \textbf{Tasks} & \textbf{Passed} & \textbf{Pass rate (\%)} \\
\midrule
Copilot CLI & 203 & 115 & 56.7 \\
Codex & 203 & 105 & 51.7 \\
\bottomrule
\end{tabular}
\end{table}

\rqanswer{Across all ten configurations, incomplete migration is the most
frequent failure label (36.5\%), followed by wrong semantic output (31.5\%)
and wrong root cause (28.6\%). A failure trajectory may receive multiple labels.
These observations motivate \emph{DepUpgrader}, which combines evidence-grounded
contract checks with repository-wide migration and validation. On the two
best-performing configurations, Copilot CLI and Codex (both with GPT-5.5),
the skill reports relative pass-rate gains of 15.0\% and 6.1\%, respectively.}

\section{Related Work}

\subsection{Dependency-Upgrade Benchmarks and Repair}

Among earlier benchmarks, BUMP is closest to our setting: it defines a breaking
dependency update as a commit that changes a dependency version and turns a successful build into
a failing one~\cite{reyes2024bump}. \benchmarkname\ extends this setting across
ecosystems, treats repair as an agent task, and covers hidden test, runtime,
lockfile, package-manager, and setup failures.

Concurrent benchmarks study complementary problems. SWE-Chain evaluates chained
release upgrades across 12 chains from nine Python packages, with each transition
inheriting the preceding codebase~\cite{lam2026swechain}; \benchmarkname\ instead
isolates independent dependency upgrades in downstream projects across five ecosystems.
DI-BENCH evaluates inference of the dependencies needed to execute a
repository~\cite{zhang2025dibench}, rather than repair after a known upgrade.

Breaking-Good analyzes breaking updates using build logs and dependency
trees~\cite{reyes2024breakinggood}. Repair approaches include LLM-based fixes for
compilation failures caused by dependency updates~\cite{reyes2026byam}, automated
repair on Java projects~\cite{fruntke2025automatically}, and upgrade workflows
with multiple steps performed by LLM agents~\cite{tawosi2025llmagents}.
UPGRADVISOR combines program analysis and
hardware tracing to assess updates~\cite{david2022upgradvisor}, while specialized
agents target dependency-related build repair~\cite{macho2026agentbased}.
\benchmarkname\ complements these methods with executable evaluation over
broader hidden upgrade failures.

\subsection{Repository-Level Benchmarks}

Repository-level benchmarks provide useful evaluation templates. SWE-bench uses
real GitHub issues, containerized environments, and fail-to-pass tests
\cite{jimenez2024swebench}; MigrationBench and FreshBrew evaluate Java runtime
migration~\cite{liu2025migrationbench,may2025freshbrew}; PyMigBench targets
Python library migration~\cite{islam2023pymigbench}; and CI-Repair-Bench focuses
on repair of CI workflows~\cite{muna2026cirepairbench}. GitTaskBench spans 54 workflow
tasks and identifies environment setup and dependency resolution as prominent
agent failure modes~\cite{ni2026gittaskbench}.

BeyondSWE includes dependency migration as a task category
\cite{chen2026beyondswe}. As discussed in Section~\ref{sec:motivation},
\benchmarkname\ addresses construction risks in this setting through
manifest/code/test decomposition, Docker-based environments, held-out tests, and a four-state
oracle that isolates dependency-upgrade repair across ecosystems.

\section{Threats to Validity}

\subsection{Construct Validity}

\benchmarkname\ measures dependency-upgrade repair using upstream project tests
and held-out test patches. Four-state oracle validation checks that resolving
the held-out failure requires both the dependency upgrade and a source-level repair. Success
therefore reflects correctness against this executable contract, whose scope
depends on the behavior covered by the tests.

\subsection{Internal Validity}

All configurations use the same packaged tasks, with pinned repository commits,
recorded container settings, and pre-release validation. External package
registries and network resources may change between runs.

Failure labels are assigned by models and may reflect shared biases or
ambiguous evidence. Independent annotation and adjudication reduce disagreements
but do not establish that every label is correct.

The source pull requests are public and may have appeared in model training
data. During evaluation, developer repairs and held-out tests are hidden,
Git history is pruned, and upstream solution access is restricted while
dependency downloads and documentation remain available. These controls
address solution retrieval during a trial.

\subsection{External Validity}

The benchmark samples merged dependency-bot pull requests with reproducible
environments and oracle-validated repairs. We focus on upgrades that require
source-level repair beyond bot-generated changes, rather than routine updates
already handled by existing dependency bots. Manually initiated, abandoned,
or non-reproducible upgrades fall outside this sampling scope.

\section{Conclusion}

We introduced \benchmarkname, a benchmark of 203 executable dependency-upgrade
repair tasks across five ecosystems, constructed from 4,660 candidate
dependency-update pull requests mined from over 400 real-world GitHub
repositories. By separating dependency changes, source
repairs, and held-out tests, its four-state oracle establishes upgrade-induced
breakage and repair necessity under each task's validation contract. Evaluation
of ten coding-agent configurations shows that this maintenance task remains
challenging: the highest observed success rate is 100/203 (49.3\%), achieved by
Copilot CLI with GPT-5.5, with performance varying across configurations and
ecosystems. Our case studies and failure analysis show that agents often
leave affected code unchanged or preserve behavior that the upgraded dependency
no longer supports. We therefore develop \emph{DepUpgrader}, a lightweight
skill that guides agents to consult migration documentation, update affected
code, and test their changes. On the two best-performing configurations,
Copilot CLI and Codex (both with GPT-5.5), the skill reports relative pass rate
gains of 15.0\% and 6.1\%, respectively.
\benchmarkname\ provides a common set of tasks for
evaluating how well coding agents handle dependency upgrades.

\section{Data Availability}

The \benchmarkname\ benchmark and its associated artifacts are available at
\url{https://github.com/Jacksadventure/DepBench}. The research data is also
archived on Zenodo: \url{https://doi.org/10.5281/zenodo.21099118}.

\clearpage
\bibliographystyle{IEEEtran}
\bibliography{references}

\end{document}